\documentclass[graybox]{svmult}

\usepackage{type1cm}        
\usepackage{makeidx}         
\usepackage{graphicx}        
\usepackage{multicol}        
\usepackage[bottom]{footmisc}

\usepackage{newtxtext}       %
\usepackage{newtxmath}       

\usepackage{todonotes}
\setuptodonotes{inline}

\usepackage{natbib}
\setcitestyle{authoryear}

\makeindex             

\usepackage{rotating}
\usepackage{tabularx}
\usepackage{etoolbox}
\AtBeginEnvironment{quote}{\itshape}
\usepackage{hyperref}

\begin{document}

\title*{Teaching Literature Reviewing for Software Engineering Research}
\author{Sebastian Baltes and Paul Ralph}
\institute{Sebastian Baltes \at University of Bayreuth, Germany, \email{sebastian.baltes@uni-bayreuth.de}
\and Paul Ralph \at Dalhousie University, Canada, \email{paulralph@dal.ca}}
%
%
\maketitle

\abstract{
The goal of this chapter is to support teachers in holistically introducing graduate students to literature reviews, with a particular focus on secondary research.
It provides an overview of the overall literature review process and the different types of literature review before diving into guidelines for selecting and conducting different types of literature review.
The chapter also provides recommendations for evaluating the quality of existing literature reviews and concludes with a summary of our learning goals and how the chapter supports teachers in addressing them.
}

\section{Introduction}
\label{sec:introduction}

Literature reviews are a crucial part of academic research.
Although this is apparent to experienced researchers, it might not be immediately clear to early-stage graduate students.
To motivate the topic, a teacher might start with a rhetorical question:

\begin{quote}
    Have you ever written a paper for a seminar, a thesis, or a research project?
    If so, you probably already have done a literature review.
\end{quote}

While students often perform ad hoc reviews as part of seminars or theses, those reviews are usually informal.
It is important to motivate that such rather unstructured \emph{ad hoc} literature reviews might be acceptable for these contexts, but rigorous scientific research requires more systematic approaches.

After introducing the topic of literature reviews, one can take a step back and talk about different ways of doing research.
In general, we can stratify scholarship and scholarly writing into \textit{primary}, \textit{secondary}, and \textit{tertiary} research.
This distinction is crucial to understanding the different types of literature review that can be performed. 

\begin{trailer}{Primary Research}\textit{Primary} research involves making observations in the broadest sense of collecting data about objects that are not studies themselves. Computing code metrics, administering questionnaires, interviewing participants, counting bugs, collecting documents, downloading source code, and taking field notes while observing a retrospective meeting are all primary research.
\end{trailer}

Ideally, different research groups study the same phenomenon independently, allowing other researchers to collect, compare, and aggregate individual results.
Since these researchers did not conduct the study themselves but reused their results, this process is called \emph{secondary research}.

\begin{trailer}{Secondary Research}\textit{Secondary} research involves analyzing, synthesizing, and critiquing primary studies. Ad hoc reviews, case surveys, critical reviews, meta-analysis, meta-synthesis, and scoping reviews are all types of secondary research. 
\end{trailer}

Secondary research is central to evidence-based practice~\citep{kitchenham2004evidence}.
On the one hand, important decisions should typically be made based on the balance of evidence rather than a single study.
On the other hand, from the perspective of a practitioner, it is infeasible to read every study on a certain topic.
Secondary research can help solve this discrepancy.
Another aspect is that in many fields, individual studies are too small to produce accurate estimates of population parameters, for example, to assess the strength of the relationship between two variables.
However, not every paper that somehow combines the results of multiple other papers is automatically of value.
Some secondary research degrades into predominately descriptive ``papers about papers'' with limited scope and usefulness.

At a certain point in time, the body of knowledge on a certain phenomenon or a group of phenomena might be large enough to enable \emph{tertiary research}.

\begin{trailer}{Tertiary Research}\textit{Tertiary} research has two related meanings: (1) analyses of groups of secondary studies (meta-reviews); (2) summaries or indices of broad areas of scientific knowledge as found in textbooks, encyclopedia entries, etc. 
\end{trailer}

Although tertiary studies can be valuable in specific circumstances, we should expect them to be rare.

In summary, literature reviews are crucial for understanding, structuring, and synthesizing collective knowledge on a certain topic, making empirical results more accessible to researchers and practitioners.
This is particularly true for topics discussed in the software engineering research community.

The purpose of this chapter is to outline how to holistically introduce students to literature reviews, with a particular focus on secondary research.
We provide an overview of the different types of literature review before diving into guidelines for selecting and conducting the appropriate literature review type depending on the research context.
We also want to support teaching students to evaluate the quality of existing literature reviews and conclude the chapter with a summary of our learning goals and how the chapter supports teachers in addressing them.

\section{Types of Literature Review}
\label{sec:types}

This section provides a broad overview of the literature review landscape, including references for further reading.
As mentioned in the introduction, the first literature review performed by students is usually \emph{ad hoc}.
Although such reviews might be appropriate for seminar papers or theses, it is nonetheless important to understand their limitations.
This section is partially based on a previously published short paper that motivates the need for more mature secondary research in software engineering~\citep{RalphBaltes2022}.
We are convinced that educating graduate students is an essential step toward achieving that goal.

\begin{trailer}{In-class Suggestion}Starting the lecture with all \emph{types of literature review} might overwhelm students.
An alternative to this bottom-up approach would be to start top-down, i.e., first outline the \emph{process} (see the in-class suggestion in Section~\ref{sec:process}) and then provide detailed information on the types.
The types themselves can be introduced starting with examples that are aligned with the groups' interests and backgrounds.
\end{trailer}

In the following, we introduce the seven types of literature review that are outlined in the above-mentioned paper.
Tables~\ref{tab:types-categorization} and \ref{tab:types-description} summarize the content of this chapter.

\begin{trailer}{Ad Hoc Review}An \textit{ad hoc} literature review is simply a discussion of unsystematically selected literature, as contained in most research papers as part of a background or related work section.
\end{trailer}

Ad hoc reviews may develop theory~\citep{ralph2018toward}, or integrate a new theory into existing literature~\citep{stol2016grounded}.
They can further support a position paper or tertiary scholarship. 
Ad hoc reviews use purposive sampling~\citep{baltes2020sampling}; that is, researchers purposefully select papers or studies that are useful, relevant, or support their arguments.
Ad hoc reviews are often appropriate, for example, to support theory development, identify promising research topics, or prepare for a comprehensive exam.
However, they also come with major limitations.

\begin{warning}{Attention}Ad hoc reviews suffer from important limitations. Their unsystematic nature introduces sampling bias and defies replication.
\end{warning}

An unsystematic \emph{ad hoc} approach may lead to cherry-picking of evidence supporting authors' arguments~\citep{gough2017introduction}.
Therefore, ad hoc reviews are inappropriate for supporting empirical statements such as ``x causes y'', ``most people/objects have property P'', or ``process P has structure S or follows rules R''.
Those limitations are best understood by contrasting ad hoc reviews with the various systematic approaches that exist, which we will introduce in the following.

Before we dive into the different types of \textit{systematic reviews}, we want to clarify the term.
Historically, its meaning differs between research communities.
The term can be used to describe a systematic review that uses meta-analysis of quantitative studies (especially experiments) to assess the strength of the evidence for specific, usually causal, propositions.
In the context of this chapter, to avoid confusion, we refer to this type of review as \textit{meta-analysis} and define \textit{systematic reviews} as follows.

\begin{trailer}{Systematic Review (Definition)}A \emph{systematic review} is a literature review that employs a systematic (hence the name), replicable process of selecting primary studies for inclusion, including case surveys, critical reviews, meta-analyses, meta-syntheses, and scoping reviews. 
\end{trailer}

This double meaning is due to the history of systematic reviews. The concept of a meta-analytic systematic review emerged from health and medicine in the late 20\textsuperscript{th} century~\citep{purssell2020brief}, when practitioners could not keep up with the accelerating production of research. When Chalmers~\citep{chalmers1993cochrane} founded the Cochrane Library in 1993,
the medical community coalesced around using meta-analysis to inform evidence-based practice. Now, \textit{systematic review} is often  conflated with \textit{meta-analysis}. However, other types of systematic reviews have also been around for decades. Scoping reviews were proposed no later than \citeyear{arksey2005scoping} \citep{arksey2005scoping}. Meta-synthesis goes back at least to \citeyear{jensen1996meta} \citep{jensen1996meta}. Case surveys were proposed as far back as \citeyear{lucas1974case} \citep{lucas1974case}. 

\clearpage

\begin{trailer}{Systematic Review (Process)}
\emph{Systematic reviews} begin by applying a search strategy to identify primary studies that meet pre-established criteria. Most types of systematic reviews seek to identify \textit{all} the primary studies that meet the selection criteria by combining various techniques to mitigate sampling bias and publication bias as part of the search strategy. Guidelines for developing search strategies are available.
\end{trailer}

The \emph{ACM SIGSOFT Empirical Standards for Software Engineering}\footnote{\href{https://www2.sigsoft.org/EmpiricalStandards/docs/?standard=SystematicReviews}{https://www2.sigsoft.org/EmpiricalStandards/docs/standards}} list various techniques to mitigate sampling bias and publication bias~\citep{ralph2020acm}, which researchers can adopt when developing their search strategy: 

\begin{itemize}
\item backward and forward snowballing searches
\item checking profiles of prolific authors in the area
\item searching both formal databases (e.g., DBLP) and indexes (e.g., Google Scholar)
\item searching for relevant dissertations
\item searching pre-print servers (e.g., arXiv)
\item soliciting unpublished manuscripts through mailing lists or social media
\item contacting known authors in the area
\end{itemize}

This systematic search yields a list of primary studies.
The way these primary studies are then analyzed determines the \textit{type} of systematic review. 
If the literature review includes gray literature such as blog posts and whitepapers, it is sometimes referred to as a \emph{multivocal literature review}~\citep{DBLP:journals/infsof/GarousiFM19}.

\subsection{Meta-analysis}
\label{sec:meta-analysis}

Although systematic reviews can be conducted for various different types of primary studies, an archetypal systematic review analyzes a set of randomized controlled experiments with the same independent and dependent variables.
\emph{Meta-analyses} are then used to aggregate the results of these primary studies.

\begin{trailer}{Meta-analysis}A \emph{meta-analysis}~\citep{glass1976primary} analyzes a set of quantitative studies, usually randomized controlled experiments, with the same independent and dependent variables to statistically aggregate the results of the primary studies into a global effect size estimate.
\end{trailer}

Consider the following scenario: Ten different experiments randomized software engineering undergraduate students into a control group (who complete tasks individually) and a treatment group (who complete tasks in pairs) to compare individual vs. pair programming. The dependent variable was the number of tasks completed successfully. Each primary study reports the results of an independent samples t-test including the mean and standard deviation for each group, the t-statistic, the p-value, Cohen's $D$ (effect size), and the 95\% confidence interval for $D$. 
Our aim is to combine the results of these ten experiments to estimate the effect of pair programming on effectiveness. Suppose that four studies found a negative effect, three found no significant effect, and three found a positive effect.
Can we conclude, based on these study results, that the effect is negative?

\begin{warning}{Attention}\textbf{Do not} use vote counting to aggregate the results of primary studies, for example, by concluding that an effect is negative because the studies reporting negative results outweigh the studies reporting positive results. 
\end{warning}

In our scenario, counting votes would mean that we concluded that the effect is negative because four negative results outweigh three positive results.
Vote counting is invalid because primary studies can have wildly different sample sizes and quality levels.
What if the studies that found positive effects were much larger and more rigorous while the studies that found negative effects were small and confounded?
What if, when the three studies without significant results are aggregated, together, their results are significant? 

Instead of vote-counting, we apply meta-analysis~\citep{glass1976primary}, that is, we statistically aggregate primary study results into a global effect size estimate. This is often possible with the summary data reported in papers, without the original datasets. 
Meta-analysis can aggregate results from other kinds of (quantitative) methods as long as the studies have the same independent and dependent variables or overlapping sets of variables. The more complicated the overlaps, the more complicated the meta-analytic model. A comprehensive tutorial on statistical procedures for meta-analysis is beyond the scope of this paper but is readily available~\citep{borenstein2021introduction}.

Meta-analytic reviews essentially have the same research question as the studies being reviewed. The purpose of the meta-analysis is to reach a more reliable and robust conclusion by aggregating all available data, implying two important criteria for meta-analysis:

\begin{enumerate}
\item researchers should go to great lengths to find \textbf{all} relevant studies
\item researchers \textbf{must} evaluate the quality of each primary study and either exclude low quality studies or include study quality as a covariate in the meta-analytic model
\end{enumerate}

In summary, when scientists equate systematic reviews with evidence-based practice, they usually mean meta-analytic reviews. Meta-analysis aggregates quantitative studies that investigate the same or overlapping hypotheses. \textit{They do not simply describe} existing research. 
Meta-analysis is rare in software engineering.
While there are several good examples \citep{shepperd2014researcher,hannay2009effectiveness,rafique2012effects}, quality meta-analysis is dwarfed by superficial scoping reviews~\citep{cruzes2011research}.

\subsection{Meta-synthesis}
\label{sec:meta-synthesis}

Since quantitative and qualitative research are both common in the software engineering research community, we want to introduce the analogue of meta-analysis for qualitative research: \emph{meta-synthesis}.

\begin{trailer}{Meta-synthesis}\emph{Meta-synthesis} refers to a family of methods of aggregating qualitative studies~\citep{jensen1996meta}. After identifying the primary studies, the researcher applies hermeneutical and dialectical analyses to understand each primary study, translate them into each other, and construct an account of the body of research; for example, a theory of the central phenomenon that unites the primary studies.
\end{trailer}

Other names for \emph{meta-synthesis} are thematic synthesis, narrative synthesis, meta-ethnography, and interpretive synthesis.
Meta-synthesis requires expertise in qualitative methods and familiarity with the underlying philosophical assumptions.
Without a deep understanding of hermeneutical~\citep[see][]{ricoeur1981hermeneutics} and dialectical~\citep[see][]{sep-hegel-dialectics} analyses, one should not attempt to perform meta-synthesis.
In principle, meta-synthesis can be applied to both qualitative and quantitative work. In practice, such combinations are philosophically strained. 

\begin{warning}{Attention}Meta-synthesis is \textbf{not} organizing papers into categories (as in scoping reviews). Meta-synthesis is the process of synthesizing a credible, nuanced account of a phenomenon from prior qualitative findings.
\end{warning}

\subsection{Case Survey}
\label{sec:case-survey}

A \emph{case survey's} primary studies are (typically qualitative) case studies in the broadest sense (i.e., a scholarly account of some events).
Experience reports and gray literature may or may not be included, depending on the study's purposes.

Unlike meta-synthesis, however, a case survey transforms qualitative accounts into a quantitative dataset that supports null-hypothesis testing.
Case surveys share the philosophy of meta-analysis (positivism), not meta-synthesis (constructivism)~\citep{bullock1987case}. 

\begin{trailer}{Case Survey}A \emph{case survey} transforms the results of (typically qualitative) case studies into a quantitative dataset that supports null-hypothesis testing.
\end{trailer}

Case surveys typically begin with a priori hypotheses and an a priori coding scheme. The researcher reads each case and extracts data into the coding scheme, often using simple dichotomous variables like `did the team have retrospective meetings? [yes/no]' or `does the case mention coordination problems? [yes/no]' The resulting dataset is often too sparse for regression modeling, so researchers use simple bivariate correlations to test hypotheses~\citep{bullock1987case}. 

The Rand Corporation proposed case studies as a ``way to aggregate existing research''~\citep{lucas1974case}, quickly picked up by Yin~\citep{yin1975using}, and later elaborated in management~\citep{bullock1987case,larsson1993case}.
Today, case surveys, or case meta-analyses, are widely used in management and information systems research~\citep{jurisch2013using}. Although software-engineering-specific case surveys are available~\citep{melegati2020case,petersen2020guidelines}, they remain rare.
However, case surveys have been used to investigate strategic pivots in software start-ups~\citep{bajwa2017failures} and how organizations select component sourcing options~\citep{petersen2017choosing}. 
Case surveys have great potential in software engineering research because case studies are so common. 

\subsection{Critical Review}
\label{sec:critical-review}

The term \textit{critical review} has different meanings in different research communities. We focus on its meaning in software engineering research.

\begin{trailer}{Critical Review}A \textit{critical review} analyzes a sample of qualitative or quantitative primary studies to support an argument or critique, often of a meta-scientific nature.
\end{trailer}

For example, Stol et al. \citep{stol2016grounded}'s critical review of the use of grounded theory in software engineering \textit{criticizes} method slurring; that is, claiming to have used a research methodology that was not actually used to create illusory legitimacy.
Similarly, Baltes and Ralph's critical review~\citep{baltes2020sampling} \textit{criticizes} how software engineering researchers often overstate sample representativeness and conflate random sampling with representative sampling. In fact, critical reviews in software engineering often investigate methodological topics such as how ethnography is reported \citep{zhang2019ethnographic} or how qualitative research is synthesized \citep{huang2018synthesizing}. 
Critical reviews differ from case surveys and meta-analyses in two important ways.

First, meta-analytic reviews aggregate evidence on causal relationships to generate evidence-based recommendations, while critical reviews critically evaluate (methodological) issues.
Critical reviews are not done to support evidence-based practice or to summarize evidence for a theory. Critical reviews are part of the \emph{meta-scientific discourse}, that is, the conversations that a scientific community has internally about how it conducts research. 

Second, for many critical reviews, including \textit{all} relevant primary studies is impossible and unnecessary. For example, a critical review of adherence to the \textit{Introduction, Method, Results and Discussion framework} (IMRaD) framework~\citep{sollaci2004introduction} could include all software engineering papers ever written. Instead, a random sample of papers from a selection of leading journals and conferences is sufficient because critical reviews do not assess causal claims, therefore publication bias---``what if significant results were published but non-significant results were not?''---is irrelevant. 

Analysis performed within a critical review can be quantitative, qualitative, or both. However, critical reviews typically adopt a \textit{critical} stance; that is, they go beyond mere description and offer specific critiques of the work being reviewed. 

\subsection{Scoping Review}
\label{sec:scoping-review}

What is often called a \emph{systematic mapping study} in software engineering~\citep{petersen2008systematic} is generally called a \emph{scoping review} elsewhere, for example, in health, medicine and psychology. 

\begin{trailer}{Scoping Review}The purpose of a \emph{scoping review} is to understand the state of research on a particular topic.
This is typically done by mapping primary studies into categories.
Therefore, they are often called \emph{systematic mapping studies} in software engineering research. 
\end{trailer}

Scoping reviews are primarily descriptive; they count the number of studies on a topic.
They often organize studies by research method, subtopic, authors, geographical location, publication venue, etc.
They often conclude that more research is needed on particular subtopics.
For example, Mohanani et al. \citep{mohanani2018cognitive} mapped primary studies according to which cognitive bias (e.g., confirmation bias) they investigated and in which area of software development (e.g., design, management) they investigated it, then called for more research on \textit{debiasing}, that is, preventing or mitigating cognitive biases.

\begin{warning}{Attention}The problem with scoping reviews is that they typically do not meet any of the core purposes of secondary research.
\end{warning}

When comparing \emph{scoping reviews} to \emph{meta-analyses} and \emph{case surveys}, one notices that the latter synthesize the results of many studies to answer specific empirical (often causal) questions about the world. Scoping reviews include a similar search, but typically do not provide sufficient quantitative synthesis to answer important empirical questions. Therefore, scoping reviews do not inform evidence-based practice as meta-analyzes and case surveys do. 
\emph{Meta-synthesis} involves deep, theory-oriented reinterpretation of related qualitative studies. Although scoping reviews can include qualitative analysis (e.g., mapping or categorization), that analysis is often too superficial to generate novel and useful theories.

\emph{Critical reviews} use a sample of papers to demonstrate an important pattern for the internal discourse of a scientific community. 
While scoping reviews often give recommendations regarding future research, they focus on an empirical topic (e.g., cognitive biases in SE); not a meta-scientific topic (e.g., construct validity); therefore, they are not configured, from the outset, to deliver useful meta-scientific critique.

In summary, scoping reviews begin like other kinds of systematic reviews but stop short of \textit{synthesizing} the data into aggregate empirical results, theory, or meta-scientific critique. This is by definition: If a scoping review applies a meta-analytic model to aggregate primary study results, or applies hermeneutics and dialectics to synthesize qualitative accounts, or develops an evidence-based critique of a scientific practice, it is no longer a scoping review; it is a meta-analysis, a case survey, a meta-synthesis, or a critical review. Therefore, some authors recommend a scoping review ``as a precursor to a systematic review''~\citep{munn2018systematic}. 

\subsection{Rapid Review}
\label{sec:rapid-review}

Before conducting a systematic literature review, one should be aware that they require a considerable time investment.
In some situations, decision-making cannot wait until this time-consuming process is finished. 

\begin{trailer}{Rapid Review}A \textit{rapid review} is a meta-analysis that makes methodological compromises to reduce the completion time~\citep{ganann2010expediting}.
\end{trailer}

Ganann et al. found many such compromises including restricting the literature search, truncating results, omitting techniques for overcoming publication bias (e.g., reference snowballing), streamlining screening and data extraction, and skipping quality assessment~\citep{ganann2010expediting}. 
Some authors argue that rapid reviews are a suitable means for practitioners to provide evidence in a timely manner~\citep{CartaxoPintoSoares2020}.
However, from a scientific point of view, rapid reviews are justified \textit{if and only if} evidence is needed to support imminent decisions, and waiting for a comprehensive meta-analytic review would be harmful.
These conditions occur in health and medicine, for example, when an unprecedented viral pandemic strikes.
These conditions do not occur frequently in software engineering.

\begin{warning}{Attention}The term \textit{rapid review} should not be used to legitimize bad systematic reviews where there is no urgent need for immediate results.
\end{warning}

In general, software-engineering-related topics rarely have so many primary studies that it would take more than a year to complete a comprehensive review.
Therefore, in scholarly research, rapid reviews should be the exception.

\begin{table}
\caption{The Seven Types of Literature Review: Categorization}
\label{tab:types-categorization}
\begin{tabular}{ l l l l l}
\hline\noalign{\smallskip}
Type & Systematic & Purpose & Primary Studies & Analysis \\
\noalign{\smallskip}\svhline\noalign{\smallskip}
Ad hoc review & no & discuss & any & any \\ 
Meta-analysis & yes & explain \& predict & quantitative & quantitative \\
Meta-synthesis & yes & explain & qualitative & qualitative \\
Case survey & yes & explain \& predict & qualitative & quantitative \\
Critical review & yes & prescribe & any & any \\
Scoping review & yes & describe & any & both \\
Rapid review & yes & explain \& predict & quantitative & quantitative \\
\noalign{\smallskip}\hline\noalign{\smallskip}
\end{tabular}
\end{table}

\begin{table}
\caption{The Seven Types of Literature Review: Comparison}
\label{tab:types-description}
 \footnotesize
\begin{tabular}{l l}
\hline\noalign{\smallskip}
Type & Approach \\
\noalign{\smallskip}\svhline\noalign{\smallskip}
Ad hoc review & discuss purposively-selected related work \\ 
Meta-analysis & estimate effect sizes by aggregating results of similar quantitative studies \\
Meta-synthesis & synthesize the findings of numerous studies using qualitative analysis \\
Case survey & test causal hypotheses by aggregating case study results \\
Critical review & defend a position and make recommendations by analyzing a sample of papers \\
Scoping review & describe an area of research and map studies into meaningful categories \\
Rapid review & a meta-analytic review that compromises rigor for speed \\
\noalign{\smallskip}\hline\noalign{\smallskip}
\end{tabular}
\end{table}

\section{Guidelines}
\label{sec:guidelines}

After the previous section introduced different types of literature review, this section provides advice for selecting, performing, and evaluating them.
We start by describing the overall literature review process (Section~\ref{sec:process}) and then continue to discuss how students can use \emph{ad hoc reviews} and \emph{scoping reviews} to screen literature (Section~\ref{sec:screening}) and how they can subsequently reflect on which secondary research method might be appropriate (Section~\ref{sec:reflecting}).
We continue with recommendations on performing secondary research (Section~\ref{sec:recommendations}) and conclude with advise on evaluating literature reviews (Section~\ref{sec:evaluating}).
Rather than comprehensive guidelines, we outline the main pitfalls as well as anti-patterns, and highlight crucial aspects.

\begin{trailer}{Learning Goals}The learning goals of a lecture on literature reviews are that students (1) understand the overall process, (2) can independently conduct \emph{ad hoc} and \emph{scoping reviews} according to our suggestions, (3) are aware of the existing secondary research methods, and (4) can evaluate existing literature reviews.
Actually performing secondary research is out of scope for such a lecture.
If a teacher wants to expand the scope of the lecture, we suggest focusing on \emph{case surveys} rather than \emph{meta-analysis} and \emph{meta-synthesis} (reasons for that are mentioned throughout this section).
\end{trailer}

\subsection{The Literature Review Process}
\label{sec:process}

\begin{figure}
\centering
\includegraphics[angle=90,width=0.65\linewidth]{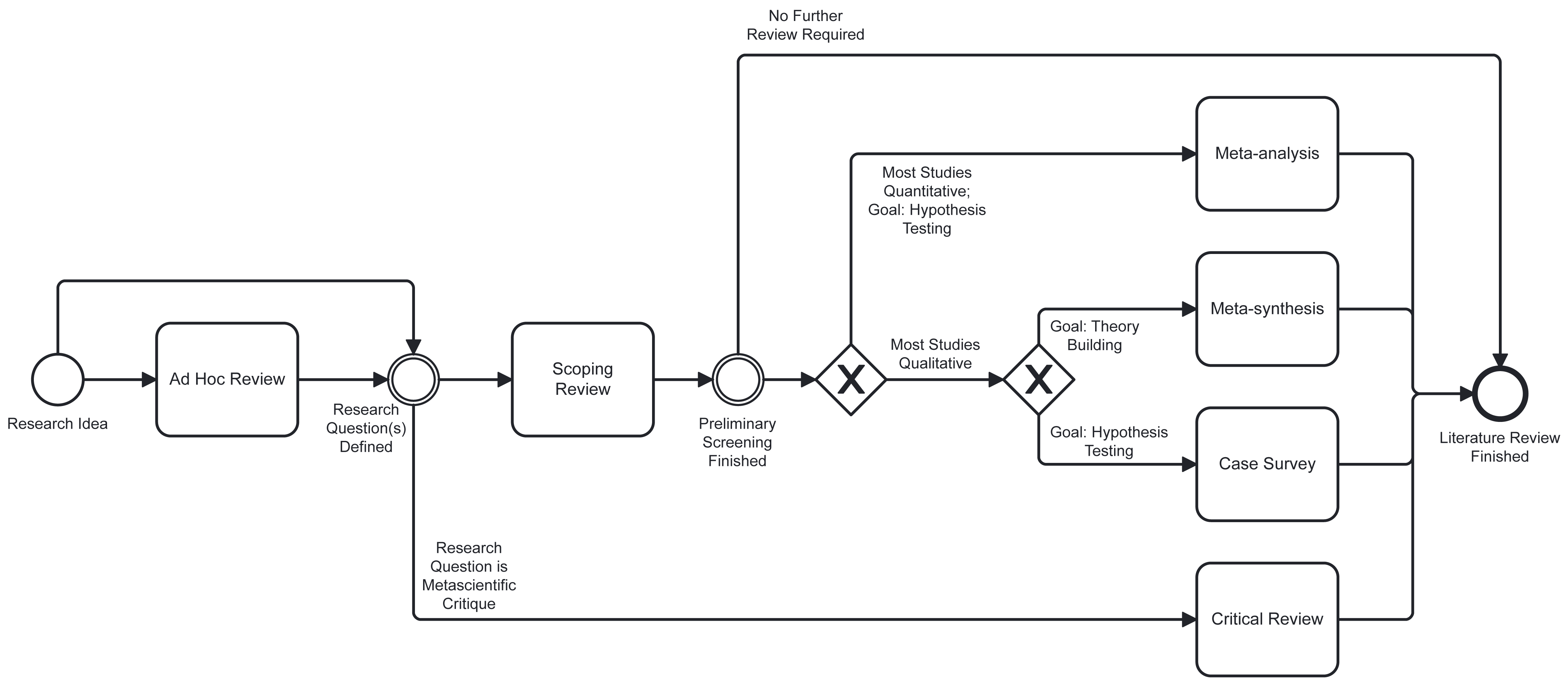}

\caption{Process of selecting literature review types at different stages of a research project.}
\label{fig:selection}       
\end{figure}

Figure~\ref{fig:selection} outlines the process of selecting appropriate literature review types at different stages of a student's research project.
In some cases, one might start with a rough research idea and then perform an \emph{ad hoc review} to assess the idea and its novelty, informally gathering a first overview of existing related work.
This information can then be used to formulate more specific research questions, which could also be defined without first conducting an ad hoc review.
Having defined specific research questions, the next step would be to perform a \emph{scoping review} to thoroughly understand the research on the topic of interest.

\begin{trailer}{In-class Suggestion}In class, the teacher can demonstrate how tools such as \href{https://scholar.google.com/}{Google Scholar} and \href{https://dblp.org/}{DBLP} can be used for \emph{ad hoc reviews}.
The teacher can then share a research idea that students are asked to explore in a \href{https://kb.wisc.edu/instructional-resources/page.php?id=104151}{\emph{buzz group}} setting.
The resulting papers are jointly collected, e.g., via \href{https://etherpad.org/}{Etherpad}.
Based on the resulting list, the teacher can discuss additional aspects of ad hoc reviews, such as selecting only peer-reviewed articles. 
\end{trailer}

After the initial screening phase is completed, a researcher might decide that no further literature review is required.
The next steps would then be to plan and conduct a novel study against the background of the identified related work.
Reasons for not conducting a \emph{scoping review} after the \emph{ad hoc review} include:

\begin{itemize}
    \item The \emph{ad hoc review} yielded existing studies that are so close to the research idea or research questions that one can either continue planning a replication study or decide that the idea is not worth pursuing.
    \item The \emph{ad hoc review} yielded recent \emph{scoping reviews} on the topic that can be used to further plan the intended study.
    \item The \emph{ad hoc review} yielded recent \emph{secondary research} on the topic that needs to be further screened.
    \item The research question is of meta-scientific nature and warrants a \emph{critical review} (see Sections~\ref{sec:critical-review} and \ref{sec:guideline-critical-review}).
\end{itemize}

It is important to explain to students that, generally, a larger research project should always involve a \emph{scoping review}.
Finding very similar related work late in the research process, limiting the novelty of an ongoing study, is very demotivating, not only for students.

\emph{Critical reviews} are a special case, because they are often performed on a random sample of papers selected from specific venues in a defined period of time.
As motivated in Section~\ref{sec:critical-review}, including all relevant papers is often infeasible for critical reviews, and therefore a scoping review is not a logical predecessor of a critical review.

If the \emph{ad hoc review} warrants performing a \emph{scoping review}, the resulting list of articles needs to be screened and the researcher needs to decide which form \emph{secondary research} is suitable (Section~\ref{sec:reflecting}) or if the scoping review itself is sufficient.
Reasons for not conducting \emph{secondary research} include:

\begin{itemize}
    \item The research methods of the identified papers are too diverse.
    \item The research questions of the identified papers are not aligned.
    \item Papers lack statistics or other information required to perform secondary research.
    \item The quantity of the identified papers is too low.
\end{itemize}

\begin{trailer}{In-class Suggestion}To underline the \emph{diversity of research methods and ways of presenting empirical studies in software engineering}, the teacher might contrast papers on human aspects of software engineering, mining software repositories, and software testing (to name a few examples).
Papers in psychology or medical research can serve as examples for disciplines with more uniform research methods and a more standardized structure for presenting empirical studies.
\end{trailer}

%

\subsection{Screening Literature}
\label{sec:screening}

In this section, we focus on the screening phase of the process outlined in Figure~\ref{fig:selection}.
What both \emph{ad hoc reviews} and \emph{scoping reviews} have in common is that they should start with a clearly formulated research idea, purpose, goal, or question (see also the chapter \emph{Research Design in Software Engineering} by Molléri and Petersen).

\begin{trailer}{In-class Suggestion}In class, the teacher might provide a counterexample, that is, a vague and ambiguous \emph{research goal}, and then refine it together with the students.
\end{trailer}

Based on the research idea or question, the next step is to define a search strategy that typically includes an initial set of keywords, inclusion/exclusion criteria, and the search engines to use.
Popular choices for computer science research are \href{https://scholar.google.com/}{Google Scholar} and \href{https://dblp.org/}{DBLP}.
\href{https://www.semanticscholar.org/}{Semantic Scholar} is another freely available search engine.
Commercial options include \href{https://www.webofscience.com/wos/}{Web of Science} and \href{https://www.scopus.com/}{Scopus}.

Some portals support \emph{backward and/or forward snowballing}.
Backward snowballing refers to analyzing the papers that an included paper cites, while forward snowballing refers to analyzing papers that cite an included paper.
Guidelines for snowballing in literature reviews~\citep{DBLP:conf/ease/Wohlin14} and for combining keyword-based portal searches with snowballing~\citep{DBLP:journals/infsof/WohlinKFM22} are available.

The inclusion and exclusion criteria define which paper the researchers are interested in.
Typical filters restrict the time span (e.g., only papers published in the last five years) or exclude articles not meeting certain standards (e.g., non-peer-reviewed articles or book chapters).
If keywords have different meaning in different contexts/disciplines, the inclusion/exclusion criteria can clarify the instances that the researchers want to consider.

\begin{trailer}{In-class Suggestion}Based on the previously defined research goal, the teacher can guide students through the definition of an initial set of \emph{keywords and inclusion/exclusion criteria}.
It is important to stress that both the keywords and the criteria can evolve in the course of a research project.
However, all changes need to be documented.
\end{trailer}

Conducting \emph{ad hoc} or \emph{scoping reviews} involves reading many paper titles, abstracts, and sections.
It should be stressed how important deep reading is for research.
Knowing when to skim over (parts of) a paper is a skill that students have to develop over time.
In the beginning, students should err on the side of reading more and skimming less, even if it slows them down.
We are well aware of the pressure to publish early and much, but this is hard to do without a deep understanding of the field.
Part of that understanding is, besides published research articles, being able to distinguish scientific evidence from opinionated articles, ``laws'', and recommendations.
Interesting examples in this regard are the \href{https://www.gartner.com/en/articles/what-s-new-in-the-2023-gartner-hype-cycle-for-emerging-technologies}{\emph{Gartner Hype Cycle}}, which claimed that software engineering was at its ``peak of inflated expectations'' in 2023, over 55 years after the first NATO Conference on Software Engineering~\citep{Nato1968}, or the consulting company McKinsey's claim that their consultants can measure software developer productivity, despite scientific research suggesting a much more nuanced view~\citep{DBLP:books/sp/SZ2019}.
Finally, students must understand that \emph{ad hoc} and \emph{scoping reviews} are usually not a study on their own, but a means of preparing themselves to perform novel studies or replications.

\begin{trailer}{Exercise Suggestion}An exercise could be to let students conduct a limited \emph{ad hoc review} at home, using the portals, keywords, and criteria discussed during the lecture.
Identifying a fixed number of papers (e.g., up to 20) and then summarizing a subset (e.g., three of them) in their own words, without repeating sentences from the abstracts, could be reasonable guardrails. Bonus exercise: include snowballing.
\end{trailer}

\subsection{Reflecting on Screened Literature}
\label{sec:reflecting}

In this section, we focus on the decision points in Figure~\ref{fig:selection}.
In Section~\ref{sec:process}, we already described in which cases one might not perform a \emph{scoping review} after an \emph{ad hoc review} or not perform \emph{secondary} research after a \emph{scoping review}.
Here, we will focus mainly on the selection of appropriate secondary research methods.

As mentioned above, \emph{critical reviews} are a special case.
They are meta-scientific in nature and analyze a sample of qualitative or quantitative primary studies to support an argument or critique (see Sections~\ref{sec:critical-review}).
They would usually not be done after a scoping review, but instead.
We do not expect students to conduct a critical review early in their careers, and therefore we only refer to our recommendations in Section~\ref{sec:guideline-critical-review}.

Assuming that the students have conducted a \emph{scoping review}, the question arises how they can decide whether the corpus of identified articles warrants secondary research.
If there is indeed a considerable body of knowledge, most studies are quantitative, and the goal is to test one or more hypotheses, the next step would be to perform \emph{meta-analysis} (see Sections~\ref{sec:meta-analysis} and \ref{sec:guideline-meta-analysis}). 
If most studies are qualitative and the goal is theory building, performing \emph{meta-synthesis} is possible (see Sections~\ref{sec:meta-synthesis}), given that the researcher has the appropriate expertise and experience (see \ref{sec:guideline-meta-synthesis}).
If most studies are case studies and the objective is to test hypotheses, students might consider conducting a \emph{case survey} (see Sections~\ref{sec:case-survey} and \ref{sec:guideline-case-survey}).

In a lecture on literature reviews, the concepts of \emph{meta-analysis} and \emph{meta-synthesis} can be presented, but we consider \emph{case surveys} to be the most useful type of secondary research for students in software engineering.
In the following section, we elaborate on this further.

\begin{trailer}{In-class Suggestion}Teachers can stress that, unfortunately, \emph{secondary research is rare in software engineering}.
In class, best-practice examples from software engineering can be shown along with examples from other fields.
\end{trailer}

\subsection{Literature Review Recommendations}
\label{sec:recommendations}

As mentioned above, performing \emph{meta-analysis} or \emph{meta-synthesis} will probably be too challenging for most graduate students.
Performing \emph{critical reviews} as a student is also rare.
We provide recommendations for all of these types of secondary research, but would like to stress that, besides conducting \emph{ad hoc reviews} and \emph{scoping reviews}, \emph{case surveys} are most appropriate for software engineering graduate students and hence teachers should focus on those three methods while outlining the others.

\begin{trailer}{In-class Suggestion}Focus on \emph{ad hoc reviews}, \emph{scoping reviews}, and \emph{case surveys} but also summarize \emph{meta-analysis} and \emph{meta-synthesis}, making clear that these are advanced secondary research methods usually out of scope for graduate students in software engineering.
\end{trailer}

\subsubsection{Meta-analysis}
\label{sec:guideline-meta-analysis}

In our experience, the overwhelming problem with \emph{meta-analysis} in software engineering is that, for a given topic, we almost never have enough similar quantitative studies to aggregate statistically. The research questions and statistics methods are usually too diverse, or essential properties such as effect sizes are missing.
Moreover, software engineering research can be performed in very diverse contexts, further limiting the comparability of studies.
Therefore, an attempt to conduct a true meta-analysis almost always devolves into a scoping review.

In addition, meta-analysis is hard to do well by one person alone.
Multiple researchers need to perform the abstract-and-title screening and then the full-text screening.
They need to double-enter all the data, which subsequently needs to be compared and aligned.
Hence, this is not something a graduate student can and should do alone.
One needs a team. 

Moreover, meta-analysis is very time-consuming and often takes more than a year. 
This results in pragmatic issues such as new papers being published in the process of performing the meta-analysis, but also during the paper review process after submission.
The dataset must be constantly kept up-to-date, and the statistical methods used in the meta-analysis should be automated so that statistics can be regularly updated.

Our specific recommendations for students (and supervisors) who nevertheless want to conduct a meta-analysis are:
\begin{itemize}
    \item If you want to conduct a meta-analysis, you need to read dedicated text books on the topic (e.g., \citep{borenstein2021introduction}). A solid background in statistics is helpful.
    \item Do not vote-count (see Section~\ref{sec:meta-analysis}). 
    \item It is possible to statistically model study quality instead of excluding low-quality studies. Refer to the empirical standards~\citep{ralph2020acm} to assess study quality.
\end{itemize}

In summary, the situation around meta-analysis in software engineering is a chicken-and-egg problem.
We are convinced that software engineering needs more meta-analysis, but, as we just motivated, this kind of secondary research is very hard to do and the success of students in that endeavor is improbable, because research and reporting quality in software engineering do not meet the standards of other disciplines such as medical research or psychology.

\subsubsection{Meta-synthesis}
\label{sec:guideline-meta-synthesis}

\emph{Meta-synthesis} requires a strong appreciation of qualitative research.
Crucial to this is understanding constructivism and interpretivism.
Without this understanding, meta-synthesis is virtually impossible.
Moreover, meta-synthesis should only be attempted by people with considerable experience in conducting qualitative studies.
Again, without such experience, meta-synthesis is virtually impossible.

Ideally, to do meta-synthesis well, researchers need to understand the differences between different qualitative research traditions (e.g., case study, ethnography, grounded theory, phenomenology, critical theory). 
Very few researchers in software engineering have such a deep methodological understanding, because the methods are rooted in other disciplines; hence, ``method slurring'' is common~\citep{stol2016grounded}.
Although evidence standards~\citep{ralph2020acm} can be used to assess study quality, assessing the quality of qualitative research can be more difficult because there are not as many obvious red flags, such as missing effect sizes. 
All this further complicates meta-synthesis.

Finally, when performing a meta-synthesis, one must organize the findings of the primary studies into reasonable narratives without forcing one's own expectations or assumptions onto the data. Auto-reflection~\citep{doi:10.1080/08873267.2003.9986934} is crucial. 
In summary, our recommendation is to only attempt meta-synthesis after:
\begin{itemize}
    \item reading \textit{many} books on qualitative methods,
    \item conducting several qualitative studies,
    \item reading at least one book on meta-synthesis (e.g., \citep{jensen1996meta}),
    \item reading \textit{many} studies from a range of qualitative research traditions.
\end{itemize}

Unfortunately, having a teacher who has done all of the above is not enough.
The primary analyst needs the expertise that comes from a combination of first-hand experience and second-hand study.
Therefore, meta-synthesis is not a great choice for most graduate students unless an experienced supervisor is \textit{very} engaged in the analysis.

\subsubsection{Case Survey}
\label{sec:guideline-case-survey}

In our opinion, \emph{case surveys} are a great option for graduate students because case surveys are more structured than \emph{meta-synthesis} but less difficult to perform than \emph{meta-analysis}.
One does not need to know as much about research methods as for a \emph{critical review}.
In our experience, case study is one of the most common research methods in software engineering, so there are a lot of good---and not so good---examples out there.

One problem is that, as mentioned in the previous section, method slurring is a problem in software engineering research, and this also applies to \emph{case studies}.
As a researcher, one needs to have clear rules to decide what is a case study (to include it) and what is not (to exclude it).
One cannot make this decision based on the presence (or absence) of the words ``case study'' alone.
There are different schools of thought on what constitutes a \emph{case study}. 

Yin~\citep{yin2009case} argued that a case must have clear evidence of triangulation across different kinds of data (e.g., interviews, online discussion threads, change logs); whereas Bullock and Tubbs~\citep{bullock1987case} included just about any account of the phenomenon of interest.
Another way is to include only articles that clearly identify a \emph{site} that the researchers visited.
This third way is good for excluding a lot of low-quality pseudo-cases but is problematic for studies of remote-first teams and historical cases. 

A problem is that, while guidelines exist~\citep{DBLP:journals/ese/RunesonH09, DBLP:books/daglib/0033235}, software engineering research does not have a history of rich description in case study research that, for example, sociology has.
Therefore, students might run into the problem that the analyzed articles do not report the required details.

\begin{trailer}{In-class Suggestion}Given the large number of case studies in software engineering research a and the existence of guidelines and examples, teachers should focus on this secondary research method and present the existing guidelines and best-practice examples.
\end{trailer}

For \emph{case surveys}, the statistics are simpler than for \emph{meta-analysis}, because the resulting data set is much more sparse and statistics is limited to fundamental methods such as correlations.
Structural equation modeling (SEM) or Partial least squares regression (PLS) are usually not possible to use for the reasons outlined above.

Our recommendation is that all data extraction should be performed by two researchers independently and then compared and aligned to ensure quality and assess reliability.
Bullock and Tubbs~\citep{bullock1987case} recommend asking all the authors of primary studies to fill out the data collection for their papers in a survey, so that they act as a second coder and can enter information they know that was not reported in the article.
However, it is not easy to motivate the authors to do this (or to contact them in the first place), especially for older papers.
It might help if not the student, but a senior supervisor that is well known in the field, sends out the invites but even then the response rate will most likely be far away from 100\%.

\subsubsection{Critical Review}
\label{sec:guideline-critical-review}

In theory, \emph{critical reviews} can be a great choice for a graduate student because doing a critical review usually means analyzing the research methods used in an area, finding all the common problems, and figuring out how to do a method better.
This can be a great prelude to conducting one's own research because it helps the student avoid all these pitfalls they identified in the \emph{critical review} and hence design better studies. 

However, \emph{critical reviews} suffer from some of the same problems as \emph{meta-analysis}---you need a team and the analysis can be time consuming.
Moreover, performing a \emph{critical review} on a method that the student has not or only rarely used themselves, is challenging.
An advantage of \emph{critical reviews} is that they do not require a comprehensive sample that includes---in the best case---all studies on a topic.
One only needs a defensible sample of studies.
Therefore, the sampling and analysis of paper does not take as long as for \emph{meta-analysis} and \emph{meta-synthesis}, and a slightly outdated sample does not affect the \emph{critical review} as much.

The biggest challenges we have encountered with critical reviews are that:

\begin{itemize}
    \item Most students (and also many supervisors) have insufficient research methods training to assess the primary studies. One reason for that is also the diversity of research methods in software engineering, as motivated above.    
    \item Reviewers may be unfamiliar with \emph{critical reviews} and therefore treat them as scoping reviews, expecting the authors to include all relevant studies. As described in Section~\ref{sec:critical-review}, this is not required for critical reviews.
    \item \emph{Critical reviews} have ethical implications (see below).
\end{itemize}

\begin{warning}{Attention}\textbf{Ethics:} It is hard to critique primary studies without publicly blaming the authors. We recommend to describe problems without citing the primary studies that exemplify those problems.
One can state explicitly that this is done because the additional credibility of pointing to examples does not justify the hurt feelings of the authors and the potential impact on the reputation of the researchers conducting the review.
A risk for us as authors is that reviewers might not follow this argument.
\end{warning}

\subsection{Evaluating Literature Reviews}
\label{sec:evaluating}

Most of the time, students will read or review literature reviews at some point during their studies.
Therefore, we want to teach them which aspects to look out for when assessing the quality of literature reviews.

Learning to assess the quality of existing research is a crucial skill for any scientist.
We cannot infer that a study is good just because it was peer reviewed and published in a reputable venue (or bad because it is a preprint or published in an outlet of ill repute).
The criteria described in this section are adapted from the \emph{ACM SIGSOFT Empirical Standards for Software Engineering}~\citep{ralph2020acm}. Rather than repeating all the criteria, we will focus on the most common red flags. 

\subsubsection{Evaluating Ad Hoc and Scoping Reviews}

An \emph{ad hoc review} is what students would report in the related work section of a research paper.
They are not a standalone publication.
Similarly, \emph{scoping reviews} are basically pilot studies on the way to conducting a real systematic review.
However, given the diversity of research methods and the issues with reporting empirical studies in software engineering, which we described above, they are often published as standalone articles in journals or full-paper conference tracks.
This is uncommon in other disciplines and will hopefully change as secondary research in software engineering matures~\citep{RalphBaltes2022}.

The most common red flag in an \emph{ad hoc review} is the vague feeling that the authors have not actually read, in detail, the papers being discussed.
When the reader is familiar with some of the reviewed works, and the authors' comments on them suggest serious misunderstandings, the reader worries that the review is superficial. 

In contrast, the authors of \emph{scoping reviews} do not necessarily read the primary studies in depth; they rigorously extract specific data from the studies. Some things that suggest low quality in a scoping review are:

\begin{itemize}
    \item The process of selecting the primary studies is not described in sufficient detail that other researchers could replicate it.
    \item There is no public data set.
    \item The authors did not apply techniques for mitigating sampling bias~\citep{baltes2020sampling} (e.g., reference snowballing).
\end{itemize}

\subsubsection{Evaluating Meta-analysis}

\emph{Meta-analysis} usually involves estimating the effect size of one or more causal relationships by creating a statistical meta-model to aggregate the results of numerous similar studies.
Two of the biggest validity threats to \emph{meta-analysis} are publication bias and the quality of the primary studies.
This leads to several common red flags:

\begin{itemize}
    \item The study does not have a statistical meta-model, or the model does not take into account differences in primary study size and quality. Sometimes, there might be no meta-model because the paper is a scoping review masquerading as meta-analysis.
    \item Insufficient attempts to find all relevant studies (e.g., overly restrictive search terms, no reference snowballing).
    \item Insufficient attempts to mitigate and asses publication bias (e.g., including pre-prints and dissertations)
    \item No discussion of inter-rater reliability for abstract/title and full-text screening.
    \item Missing or incomplete replication package, including data set.
\end{itemize}

For a really quick and dirty assessment, if the meta-analysis paper does not have a PRISMA diagram\footnote{\url{https://www.prisma-statement.org/prisma-2020-flow-diagram}} (showing study selection) or a funnel plot~\citep{doi:10.1177/1536867X0400400204} (visualizing publication bias), it is probably of low quality. 

\subsubsection{Evaluating Meta-synthesis}

The synthesis of qualitative research does not involve complicated statistical models, but it does involve a complicated process of comparing, contrasting, and translating concepts and findings between studies. Qualitative studies often use (and invent) different concepts to describe and explain related phenomena, so synthesizing the findings involves retelling one paper's narratives in the languages of other papers.
This is much more difficult than it sounds, so a good meta-synthesis explains---\textit{in vivid detail}---how the primary studies were coded, compared, contrasted, translated, and (eventually) integrated into themes. Red flags include:

\begin{itemize}
    \item The themes are not clearly defined, vividly explained, and grounded in \textit{many} quotations from the primary studies,
    \item The data analysis process is only briefly or superficially explained.
\end{itemize}

\subsubsection{Evaluating Case Surveys}

\emph{Case surveys} are quite similar to \emph{meta-analysis} except that a case survey's statistical meta-model is much simpler.
Each case study has the same size ($n=1$), so we do not model study size.
Since we are only concerned with the facts of the case (not the interpretations), we do not exclude low-quality cases or model study quality unless we have reason to believe a case was fabricated.
However, there are red flags to watch out for, including the following:

\begin{itemize}
    \item Insufficient attempts to find all relevant studies (e.g., overly restrictive search terms, no reference snowballing).
    \item No discussion of inter-rater reliability for abstract and full-text screening; no discussion of quality control during data extraction.
    \item Insufficient a priori justification of hypotheses. 
    \item Missing or incomplete replication package, including data set.
\end{itemize}

About that last red flag: case surveys lend themselves to \emph{HARKing} (Hypothesizing After Results are Known).
In an ideal world, case surveys, including all of their hypotheses, would be publicly pre-registered such that HARKing is easily identifiable and thus discouraged.
Without pre-registration, however, we can only look for how well each hypothesis is justified conceptually, or based on previous research.

\subsubsection{Evaluating Critical Reviews}

At risk of repeating ourselves, \emph{critical reviews} do not need to include every study that meets the selection criteria.
Critical reviews often just take a (stratified) random sample of papers from some venues the authors consider reputable, and---for the scope of a \emph{critical review}---that is fine, because the point of a critical review is to make some argument about the way research is done in some area, not to estimate the effect size of a causal relationship.
Publication bias is usually irrelevant; therefore, critical reviews do not need PRISMA diagrams or funnel plots.
Critical reviews do not exclude low-quality studies because the whole point is to critique the quality of the primary studies.

The quality of a critical review is all about the critique itself.
A good critical review identifies several problems with existing research and suggests specific ways to address these problems.
Common red flags include:

\begin{itemize}
    \item The critique seems superficial.
    \item The suggested solutions are intractable.
    \item The authors of the critical review do not seem to be experts in the method they are critiquing.
\end{itemize}

An example of the last point: a critical review of experimental designs mixing up quasi-experiments with true experiments would be a serious red flag.

\section{Conclusion}
\label{sec:conclusion}

As described in Section~\ref{sec:guidelines}, the learning goals of a graduate lecture on literature reviews should be that students:

\begin{enumerate}
    \item understand the overall literature review process and can reproduce Figure~\ref{fig:selection} capturing that process,
    \item can independently conduct \emph{ad hoc} and \emph{scoping reviews} according to our suggestions in Sections~\ref{sec:types} and \ref{sec:guidelines},
    \item are aware of the secondary research methods, that is, they can reproduce the description of \emph{meta-analysis}, \emph{meta-synthesis}, and \emph{case survey},
    \item can evaluate existing literature reviews according to the recommendations in Section~\ref{sec:evaluating}.
\end{enumerate}

As mentioned earlier, if a teacher wants to expand the scope of the lecture, we suggest focusing on \emph{case surveys} rather than \emph{meta-analysis} and \emph{meta-synthesis} because case studies are common in software engineering research and the barriers to performing that kind of secondary research are lower for students.
We hope that this chapter, together with the in-class and exercise suggestions, will support teachers in preparing a lecture on literature reviews.

\bibliographystyle{spbasic}      
\bibliography{literature}   

\end{document}